\RequirePackage{rotating}
\documentclass[manuscript,screen]{acmart}

\usepackage{epigraph}
\usepackage{ifthen}
\usepackage{url}
\usepackage{enumitem}
\usepackage{balance}
\usepackage{ulem}

\def\BibTeX{{\rm B\kern-.05em{\sc i\kern-.025em b}\kern-.08emT\kern-.1667em\lower.7ex\hbox{E}\kern-.125emX}}

\begin{document}

\title{Open Source at a Crossroads: The Future of Licensing Driven by Monetization}

\author{Raula Gaikovina Kula}
\affiliation{%
 \institution{The University of Osaka}
 \country{Japan}}
 \email{raula-k@ist.osaka-u.ac.jp}

\author{Brittany Reid}
\affiliation{%
  \institution{Nara Institute of Science and Technology}
  \country{Japan}
}
\email{brittany.reid@naist.ac.jp}

\author{Christoph Treude}
\affiliation{%
  \institution{Singapore Management University}
  \country{Singapore}
}
\email{ctreude@smu.edu.sg}
se

\renewcommand{\shortauthors}{Kula, et al.}

\begin{abstract}
The widespread adoption of open source libraries and frameworks can be attributed to their licensing. Open Source Software Licenses (OSS licenses) ensure that software can be sold or distributed as part of aggregate programs from various sources without requiring a royalty or fee. The quality of such code rivals that of commercial software, with open source libraries forming large parts of the supply chain for critical commercial systems in industry. Despite this, most open source projects rely on volunteer contributions, and unpaid library maintainers face significant pressure to sustain their projects. One potential solution for these projects is to change their licensing to ensure that maintainers are compensated accordingly for their work.
In this paper, we explore the potential of licensing to help alleviate funding issues, with a review of three different cases where OSS licenses were modified to allow for monetization. In addition, we explore licensing concerns related to the emergence of the use of artificial intelligence  (AI) in software development.
We argue that open source is at a crossroads, with a growing need to redefine its licensing models and support communities and critical software. We identify specific research opportunities and conclude with a research agenda comprising a series of research questions to guide future studies in this area.
\end{abstract}

%

\maketitle

\section{Introduction}

The open source movement has dramatically reshaped the landscape of software development, as evidenced by the massive scale of GitHub in recent years, with over 207 million public repositories.\footnote{\url{https://github.com/search/advanced}} According to the 2024 Synopsys risk analysis report~\cite{report2}, open source components and libraries form the backbone of nearly every application in every industry. 
The latest GitHub report in 2023 shows that more than 94 million developers are on GitHub, and open source, with big-name open source projects like Docker and Kubernetes, is now the foundation of more than 90\% of the world’s software~\cite{githubstate}.

With the widespread adoption of open source libraries and third-party libraries, developers have access to high-quality, reusable code that accelerates innovation and drives efficiency in application development. This adoption can largely be credited to the nature of open source licensing, which permits the free redistribution and modification of software, enabling it to be sold or distributed as part of an aggregate software package without the imposition of royalties or fees. Such licenses have not only democratized software development but also established these libraries as critical components in the supply chains of large-scale industrial systems.

The sustainability risks of these projects are no secret~\cite{githubsustain}, with GitHub openly \textit{supporting maintainers in cultivating vital, productive communities}. The majority of open source libraries rely on the altruistic contributions of volunteer developers, many of whom maintain significant software projects without financial compensation. 

One example is polyfill,\footnote{\url{https://polyfill.io/}} a popular open source library for supporting older browsers, embedded in over 100K sites using the cdn.polyfill.io domain. Due to their financial situation, the library's maintainer was unable to maintain the library. The domain and GitHub account were then purchased by another company -- only to be caught injecting malware into mobile devices through any site that embeds cdn.polyfill.io.\footnote{\url{https://sansec.io/research/polyfill-supply-chain-attack}} This supply chain attack sparked a response from the original owner on social media:

\epigraph{
{\textit{``\textit{If your website uses http://polyfill.io, remove it IMMEDIATELY.
I created the polyfill service project, but I have never owned the domain name and I have had no influence over its sale.}''}}}
{--\textit{
\url{https://x.com/triblondon/status/1761852117579427975}
}}

Another famous case arose when the maintainer of the npm libraries colors and faker sabotaged their libraries to protest a lack of financial compensation -- turning these libraries into protestware~\cite{kula2022war}. Before injection of malicious code, the maintainer expressed the intention of no longer supporting large companies with their ``free work'' and that businesses should either fork the developer’s projects or compensate them.

Maintaining OSS projects places a considerable burden on volunteers, who must balance these responsibilities with personal and professional commitments:

\epigraph{
{\textit{``\textit{Over the years, I have received many proposals to monetize this extension so I think I'll just start posting them here for fun (but not for profit). The main reason I continue to maintain this extension is because I can hardly trust others to not fall for one of these offers. I'm fortunate to have a job that pays well enough to allow me to keep my moral compass and ignore all of these propositions. I realize that not everyone has the same financial security so hopefully this thread would shed some light on what kind of pressure is put on extension developers.}''}}}
{--\textit{
\url{https://github.com/extesy/hoverzoom/discussions/670}
}}

Prior work has investigated how Open Source communities source financial incentives. 
For instance, bounties are frequently used to attract developers and motivate them to complete diverse software development tasks, such as fixing software vulnerabilities~\cite{finifter2013empirical} and bug reports~\cite{hata2017understanding}.
Krishnamurthy and Tripathi~\cite{krishnamurthy2006bounty} provided an overview of bounty programs in OSS and suggested that bounty hunters' responses are related to their workload.
Finifter et al.~\cite{finifter2013empirical} analyzed vulnerability rewards programs for
Chrome and Firefox and reported that such programs are economically effective, compared to the cost of hiring full-time security researchers.
Maillart et al.~\cite{maillart2017given} recommended that project managers should dynamically adjust the value
of rewards according to the market situation.
Zhou et al.~\cite{zhou2020studying} studied the association between bounties and the issue-addressing likelihood of
GitHub issue reports.
In the context of the question-answering process, Zhou et al.~\cite{zhou2020bounties} observed that questions
are likely to attract more traffic after receiving a bounty.
More recent studies~\cite{shimada2022github, zhou2022studying} investigate GitHub Sponsors, a service that allows GitHub users to make and accept donations.
Specifically, Shimada et al.~\cite{shimada2022github} found that sponsored developers are more active than non-sponsored developers. 

All of these financial incentives are small-scale and often not sufficient to address the sustainability concerns in open source. An emerging solution to this challenge is the reconsideration of licensing models. By altering their licensing terms, some projects aim to secure a financial return for their maintainers, thus providing an incentive for sustained project maintenance and development. 

In this paper, we review three cases of open source projects that have transitioned to different licensing models in response to financial pressures. These cases highlight the complexities and consequences of such transitions, both for the projects themselves and for the broader ecosystem of users and contributors.
We argue that the open source community is at a crucial crossroads. As the landscape of software development continues to evolve, there is a growing need to rethink how open source licenses are structured and how these projects are funded. In addressing these challenges, we outline a series of research questions aimed at exploring the future of open source licensing and funding. This research agenda is designed to spark further investigation and discussion, guiding stakeholders toward sustainable solutions that support the continued growth and health of the open source ecosystem.

\section{Motivating Cases}
In this section, we review three cases where the licensing has been modified, resulting in financial implications for developers. 
It is important to understand the different types of open source licenses approved by the Open Source Initiative (OSI).\footnote{\url{https://opensource.org/licenses}}
The two main categories of open source licenses are permissive and restrictive. Permissive licenses, such as the MIT License, allow users to use, modify, and distribute the code with few restrictions. Restrictive licenses, such as the GPL License, have more stringent requirements, such as mandating that any changes to the code be released under the same license. There are many different types of open source licenses, each with its own set of terms and conditions for using, modifying, and distributing open source software.

\subsection{Case 1: The GPL License (The Case of MySQL)}
According to Wikipedia,\footnote{\url{https://en.wikipedia.org/wiki/MySQL}} MySQL is free and open source software under the terms of the GNU General Public License and is also available under various proprietary licenses. 
In 2010, when Oracle acquired Sun, Widenius forked the open source MySQL project to create MariaDB.
The MySQL project illustrates how developers can monetize modifications to open source software. The key issue is not whether the software is free, but whether modifications adhere to a GPL-compatible license. 
Such a license would enable customers to further modify and redistribute the software.
The intent may be to make life harder for other businesses trying to profit from MySQL, such as AWS with its Aurora MySQL-compatible system or MariaDB with its MySQL fork, but it also creates barriers for engineers who want to examine the code in development to find easier alternatives to a monolithic Oracle download.

Considering more technical information related to the different use cases and how they apply to users of MySQL: 
\begin{itemize}
    \item \textit{MySQL Commercial License:} Oracle provides its MySQL database server and MySQL Client Libraries under a dual license model designed to meet the development and distribution needs of both commercial distributors (such as OEMs, ISVs, and VARs) and open source projects.\footnote{\url{https://www.mysql.com/about/legal/licensing/oem/}}
    \item \textit{For Distributors of Commercial Applications:} OEMs (Original Equipment Manufacturers), ISVs (Independent Software Vendors), VARs (Value Added Resellers), and other distributors that combine and distribute commercially licensed software with MySQL software and do not wish to distribute the source code for the commercially licensed software under version 2 of the GNU General Public License (the ``GPL'') must enter into a commercial license agreement with Oracle.
    \item \textit{For Open Source Projects and Other Developers of Open Source Applications:} For developers of Free Open Source Software (``FOSS'') applications under the GPL that want to combine and distribute those FOSS applications with MySQL software, Oracle’s MySQL open source software licensed under the GPL is the best option.
    \item \textit{For Developers and Distributors of Open Source Software under a FOSS License Other Than the GPL:} Oracle makes its GPL-licensed MySQL Client Libraries available under a FOSS Exception that enables use of those MySQL Client Libraries under certain conditions without causing the entire derivative work to be subject to the GPL.
\end{itemize}

Hence, there are cases where payment is required to use this open source project. Distributing MySQL with a non-open source project requires a commercial license. Obtaining Oracle support for MySQL also requires a commercial license. Using MySQL tools exclusively licensed to Oracle support customers, such as the MySQL Enterprise Monitor, Enterprise Backup, and various plugins, requires an Oracle support contract, which in turn requires a commercial license. Additionally, modifying MySQL source code and distributing the modifications as a non-open source offering would also require a commercial license.

On the other hand, MariaDB\footnote{\url{https://github.com/MariaDB/server}} is a community-developed, commercially supported fork of the MySQL relational database management system (RDBMS), intended to remain free and open source software under the GNU General Public License. 
The MariaDB Foundation\footnote{\url{https://mariadb.org/}} receives significant funding from the MariaDB Corporation, and much of the development is done within the corporation.

\subsection{Case 2: The BSL License (The Case of Terraform)}

The second example is a more recent case of using the BSL License, which allows for business usage. The Business Source License (BSL) represents a middle ground between open source and end-user licenses. The BSL (also sometimes abbreviated as BUSL) is considered a source-available license, meaning anyone can view or use the licensed code for internal or testing purposes, but there are limitations on commercial use.
Unlike open source licenses, the BSL prohibits the licensed code from being used in production without explicit approval from the licensor.
However, similar to open source licenses, BSL-licensed source code is publicly available, and anyone is free to use, modify, and/or copy it for non-production purposes.
The license is best described as source-available, where the source code is available for everyone to read, but there are restrictions on its use for certain companies or commercial use cases.
The Business Source License (BSL) is a recent source-available license that is gaining popularity among some previously open source companies. It has roughly the following properties:

\begin{itemize}
    \item The source code is publicly available.
    \item Use of the software is free for some use cases but requires a commercial license for others.
    \item BSL-licensed code reverts to an open source license after a certain period (e.g., 4 years). After a set period, either four years or an earlier period set by the licensor, the BSL automatically converts to an open source license of the licensor's choosing. However, the open source license must be compatible with GPL, and it usually applies only to specific software versions on a rolling basis, based on the date of release.
\end{itemize}

Famous projects using it include HashiCorp’s suite of products (Terraform, Vagrant, etc.), Couchbase, CockroachDB, Sentry, and MariaDB.

As an example, we consider Terraform,\footnote{\url{https://www.hashicorp.com/products/terraform}} a platform that provides organizations with a single workflow to provision their cloud, private data center, and SaaS infrastructure and continuously manage the infrastructure throughout its lifecycle.
Terraform recently announced its decision to transition from an open source Mozilla Public License (MPL) to a Business Source License (BSL), sparking a debate within the developer and DevOps community.

In response, an open source fork of Terraform has emerged, called OpenTofu.\footnote{\url{https://opentofu.org/}} Previously named OpenTF, OpenTofu is a fork of Terraform that is open source, community-driven, and managed by the Linux Foundation. It is still very early to forecast the impact of the change; however, early reports indicate some friction between the forks,\footnote{\url{https://opentofu.org/blog/our-response-to-hashicorps-cease-and-desist/}} with claims of copyright infringement.

\subsection{Case 3: The SSPL License (The Case of Elasticsearch)}

In the final case, we explore dual licensing that is not recognized as open source. In short, the Server Side Public License (SSPL) is not recognized as free software by the Open Source Initiative (OSI), Red Hat, and Debian, as it is discriminatory towards specific fields of use. Specifically, it discriminates against users of the software who use proprietary software within their stack, as the license requires the open sourcing of every part interacting with the service, which under these circumstances might not be possible. 

In 2012, Elastic\footnote{\url{https://www.elastic.co/}} decided to move from the open source Apache 2.0-licensed source code in Elasticsearch and Kibana to a dual license under the Elastic License and Server Side Public License (SSPL). As a dual-licensing option, the project allowed users to choose which license to apply (Elastic License or SSPL). According to Elastic, the license change ensured that users had free and open access to use, modify, redistribute, and collaborate on the code. 

It also protects their continued investment in developing products that they distribute for free and in the open by restricting cloud service providers from offering Elasticsearch and Kibana as a service without contributing back. This will apply to all maintained branches of these two products starting with the 7.11 release. Their default distribution will continue to be under the Elastic License.
The license originates from MongoDB Inc. in 2018 and is based on the GPL3 license with an additional clause that stops other companies from using the software as if they owned the product directly. 
This led to issues with Amazon Web Services,\footnote{\url{https://www.elastic.co/blog/why-license-change-aws}} which ended with an agreement on both sides. 
In the end, both Elastic and AWS are now working together to make Elasticsearch a service on AWS.

\section{Open Source in the Era of AI}

Over the last few years, we are witnessing an explosion of innovation in open source large generative AI as an alternative to platforms such as OpenAI/Microsoft, Google, or Anthropic. Meta has emerged as a leading force in open source generative AI with the release of the Llama models, alongside other significant players such as Alibaba. Potential funding from companies like Mistral has raised billions in venture funding, enterprise platforms like Databricks or Snowflake are pushing open source models, and there is a growing number of open source generative AI releases on a weekly basis. DeepSeek, a recent entrant in this space, has released its DeepSeek-R1 model under the permissive MIT License, further demonstrating how open-source AI models can be freely used and commercialized without restrictions.

While the momentum in open source generative AI is strong, a more detailed analysis shows a different reality. There are reports that open source generative AI is facing a massive funding issue.\footnote{\url{https://www.coindesk.com/opinion/2024/06/11/funding-open-source-generative-ai-with-crypto/}} When it comes to large foundation models, only large companies such as Databricks, Snowflake, Meta, or well-funded startups like Mistral are keeping up with the performance of large closed models. Most of the releases from other labs, like Databricks and Snowflake, are focused on optimized enterprise workloads, while most recent open source research is focusing on complementary techniques rather than new models. Researchers also appreciate the value of these AI models being open source, and it seems that the open source community is adapting to these changes.


As such, the open source community has been working on draft definitions and licensing on how AI will impact licensing. Recently, the Open Source AI definition has been released (1.0): 

\epigraph{
{\textit{``\textit{An Open Source AI is an AI system made available under terms and in a way that grant the freedoms to:
\\
- Use the system for any purpose and without having to ask for permission.\\
- Study how the system works and inspect its components.\\
- Modify the system for any purpose, including to change its output.\\
- Share the system for others to use with or without modifications, for any purpose.}''}}}
{--\textit{
\url{https://opensource.org/ai/open-source-ai-definition}
}}

Still, the impact is not yet known. 
There is already pushback from open source communities, with two established communities voicing their concerns. 
The first is NetBSD,\footnote{\url{https://www.netbsd.org/}} which is a Unix-like open source operating system. It is available for a wide range of platforms, from large-scale servers and powerful desktop systems to handheld and embedded devices. They have recently published a policy regarding any open source contributions:

\epigraph{
{\textit{``\textit{New development policy: code generated by a large language model or similar technology (e.g., ChatGPT, GitHub Copilot) is presumed to be tainted (i.e., of unclear copyright, not fitting NetBSD's licensing goals) and cannot be committed to NetBSD.}''}}}
{--\textit{
\url{https://mastodon.sdf.org/@netbsd/112446618914747900}
}}

This sentiment is also shared by the Gentoo\footnote{\url{https://www.gentoo.org/}} community. 

\epigraph{
{\textit{``\textit{It is expressly forbidden to contribute to Gentoo any content that has been created with the assistance of Natural Language Processing artificial intelligence tools. This motion can be revisited, should a case be made on such a tool that does not pose copyright, ethical and quality concerns.}''}}}
{--\textit{
\url{https://www.osnews.com/story/139444/gentoo-bands-use-of-ai-tools/}
}}

In summary, the landscape of open source generative AI is rapidly evolving, driven by both innovation and significant funding challenges. As major players like Meta, Databricks, and Snowflake push the boundaries of what is possible, the open source community is forced to adapt to new realities, including the need for sustainable funding models. At the same time, concerns about the ethical and legal implications of AI-generated content are prompting stricter policies from established open source communities like NetBSD and Gentoo. As we navigate these changes, finding the right balance between openness, innovation, and financial sustainability will be crucial. 

\section{Research Agenda}
In this section, we present our research agenda.
We have reviewed three cases where licensing changes in open source software were made due to funding motivations. 
These cases form initial evidence on the current state of practice, and how the nature of open source software and the funding situation will affect how open source software will be defined in the future. 

We have anecdotal evidence from blogs and developer discussions on the topic, yet there have been very few research studies that articulate the impact of lack of funding and how licensing relieves these pressures. From these motivating cases, the following questions about the possible implications have emerged. In turn, we suggest research questions that could form the basis for future research projects.  

\subsection{Initiatives to Attract Funding}

The first set of research questions refers to how changes to licenses may help increase the funding situation of open source projects. Understanding the impact of various licensing models on funding is crucial, as it can influence the sustainability and growth of these projects. By exploring different funding strategies and their effectiveness, we can identify best practices that open source projects can adopt to enhance their financial support.
\begin{enumerate}
    \item What impact do different open source licenses have on attracting funding from various sources such as individual contributors, corporations, and commercial entities?
    \item What models of funding have the most promise in supporting open source projects with restrictive licenses compared to those with permissive licenses?
    \item How can hybrid funding models (e.g., combining crowdfunding with institutional funding) be optimized to support open source projects?
    \item What are the most effective strategies for open source projects to demonstrate value to potential funders?
\end{enumerate}

\subsection{Open Source Philosophy and Financial Freedom}

The following research questions explore the delicate balance between trying to keep a project open source while obtaining financial support for project sustainability. This involves examining how different licensing models can impact project growth, developer perceptions, and the ability to form effective partnerships with commercial entities. By understanding these dynamics, we can better appreciate the trade-offs and opportunities that different licensing strategies present for maintaining the open source ethos while ensuring financial viability.

\begin{enumerate}[resume]
    \item How do developers perceive the trade-offs between more restrictive licensing (e.g., GPL) and more permissive licensing (e.g., MIT) in terms of funding opportunities and project growth?
    \item How effective are partnership models between open source projects and commercial entities in ensuring project sustainability without compromising the principles of open source?
    \item What challenges and successes have projects faced when transitioning licensing models to balance funding and open collaboration?
    \item What is the impact of generative AI on the licensing and earning potential of open source projects?
    \item How do different licensing models impact the ability of open source projects to attract and retain contributors?
    \item What role do ethical considerations play in the decision-making process for licensing changes in open source projects?
\end{enumerate}

\subsection{Impact at the Software Ecosystem Level}

The final set of research questions targets the overall disruption that will be caused by license changes. This is in terms of the rest of the software ecosystem and the supply chains that are affected. Licensing changes can have wide-ranging effects on software interoperability, innovation rates, and the commercial adoption of open source software. Understanding these impacts is essential for developing strategies that mitigate potential negative consequences while leveraging the benefits of licensing modifications.

\begin{enumerate}[resume]
    \item How do licensing choices affect the long-term sustainability of open source projects in terms of both community engagement and the security risks of supply chains?
    \item What role do community-driven funding models (e.g., crowdfunding, GitHub Sponsors) play in projects under different licensing schemes?
    \item What are the challenges and successes experienced by projects that have transitioned their licensing model in an effort to better balance funding and open collaboration?
    \item How do licensing changes influence the interoperability of open source software with other software systems?
    \item What is the impact of licensing changes on the innovation rate within the open source community?
    \item How do different licensing models affect the commercialization potential and market adoption of open source software?
\end{enumerate}
.

\section{Discussion and Conclusion}

In today’s landscape, open source is at a crucial junction, challenged but invigorated by the evolving dynamics of funding and licensing. We have explored how these elements interact, impacting developers, businesses, and the broader ecosystem. As we look towards the future, it becomes increasingly clear that sustainability in open source is not just about financial support -- it is also about fostering a legal and ethical framework that nurtures innovation and collaboration. Conversations around licensing models, whether permissive or protective, underscore the critical need for adaptability and thoughtful consideration in policy making and community engagement.
Moreover, the potential of novel funding models offers a promising horizon. These models not only ensure that projects remain viable and developers are compensated, but also maintain the ethos of openness and accessibility that is fundamental to the open source spirit.
Navigating this balance will not be without its challenges. It requires a concerted effort from all stakeholders involved -- developers, corporations, and legal experts -- to dialogue continuously, innovate funding mechanisms, and refine licensing practices. Only through such collaborative endeavors can we ensure that open source remains a powerful engine for technological advancement and a beacon for collective, creative problem-solving.
The emergence of artificial intelligence and its widespread adoption in software development add even more complexity to the conversation. 
The open source community and its commercial counterparts will need to explore how to tap into this potential market. 

To address the research questions posed, we need a systematic research approach. Confirming the anecdotal evidence requires methods to define and detect different forms of funding and licensing models. Our hope is that answering these questions will help us understand how to sustain and build resilient open source projects in an era of rapid technological advancement.

\bibliographystyle{ACM-Reference-Format}
\bibliography{2030-licence}

\end{document}